\documentclass[5p, number, sort&compress]{elsarticle}
\usepackage[]{fullpage}
\usepackage[utf8]{inputenc}
\usepackage[T1]{fontenc}
\usepackage{fixltx2e}
\usepackage{graphicx}
\usepackage{grffile}
\usepackage{longtable}
\usepackage{wrapfig}
\usepackage{rotating}
\usepackage[normalem]{ulem}
\usepackage{amsmath}
\usepackage{textcomp}
\usepackage{amssymb}
\usepackage{capt-of}
\usepackage{hyperref}
\usepackage{caption}
\usepackage{subcaption}
\usepackage{gensymb}
\usepackage{amsmath}
\usepackage{mathtools}
\usepackage[table]{xcolor}

\author[1]{D. Kim
\fnref{fn1}}
\author[1]{Z. Hao
\fnref{fn1}}
\author[2]{A. R. Mohazab
\fnref{fn1}}
\author[1]{A. Ansari\corref{cor1}%
\fnref{fn1}}
\address[1]{School of Electrical and Computer Engineering, Georgia Institute of Technology, Atlanta, Georgia, USA}
\address[2]{The Foundation for the Advancement of Sciences, Humanities, Engineering, and Arts, Vancouver, BC, Canada}
\cortext[cor1]{Corresponding author \newline Email address: azadeh.ansari@ece.gatech.edu (A. Ansari)}

\date{\today}
\title{On the Forward and Backward Motion of Milli-Bristle-Bots}
\hypersetup{
 pdfauthor={ARM},
 pdftitle={On the Forward and Backward Motion of Milli-Bristle-Bots},
 pdfkeywords={},
 pdfsubject={},
 pdfcreator={Emacs 25.1.1 (Org mode 8.3.1)}, 
 pdflang={English}}
\begin{document}

\begin{abstract}
  This works presents the theoretical analysis and experimental observations of bidirectional motion of a millimeter-scale bristle robot (milli-bristle-bot) with an on-board piezoelectric actuator. First, the theory of the motion, based on the dry-friction model, is developed and the frequency regions of the forward and backward motion, along with resonant frequencies of the system are predicted. Secondly, milli-bristle-bots with two different bristle tilt angles are fabricated, and their bidirectional motions are experimentally investigated. The dependency of the robot speed on the actuation frequency is studied, which reveals two distinct frequency regions for the forward and backward motions that well matches our theoretical predictions. Furthermore, the dependencies of the resonance frequency and robot speed on the bristle tilt angle are experimentally studied and tied to the theoretical model. This work marks the first demonstration of bidirectional motion at millimeter-scales, achieved for bristle-bots with a single on-board actuator.  
\end{abstract}

\maketitle
\section{Introduction}
Bristle robots are known for their relatively simple structure (i.e. no moving parts), simple actuation method, and their high speed of locomotion \cite{cicconofri2016inversion}. While applications such as pipe inspection \cite{becker2014spy,wang2008bristle} have been proposed, bristle-bots also serve as excellent models for studying swarm and collective behavior \cite{giomi2013swarming,klingner2014stick}, as well as animal locomotion \cite{desimone2012crawling}. From a purely theoretical perspective, the nonlinearities of their equations of motion  make them an interesting subject of study \cite{cicconofri2015motility}.

The steerable bristle bots reported in the literature are in the macro scales (>few cm) and use differential configurations or multiple actuators/motors to enable steering capabilities \cite{ioi1999mobile,schulke2011worm,Senyutkin1977bristle}. However, having multiple actuators complicates the down-scaling of the bristle-bots. Therefore, achieving bidirectionality and later full steering with a single piezoelectric actuator through careful design of the resonance modes of the bristles is a promising solution for robotics actuation in the milli/micro-scales.    
 
%The lack of restriction of axis of travel can be used for numerous applications such as pipe inspection \cite{becker2014spy,wang2008bristle}and toy models for study of their collective behaviors . 
% They have been mostly regarded as unidirectional locomotion systems \cite{cicconofri2015motility,gidoni2014crawling,becker2015approach,becker2014mechanics}, where multiple actuators are required for steering \cite{ioi1999mobile,schulke2011worm,Senyutkin1977bristle}. A simpler method enables multi-directional steering in bristle-bots by changing the actuation frequency. \cite{cicconofri2016inversion,cicconofri2015motility,desimone2012crawling,Gandra2019}. 
%Macro-scale (>cm-scale) bristle-bots have been widely used for XX and yy; however, achieving directionality and steering capability has proven to be challenging, while using a single on-board actuator. Differential configurations or multiple actuators/motors have been used to enable steering capabilities to such bristle robots [ref]. However, having multiple actuators complicates the size down-scaling of the bristle-bots. Therefore, achieving directionality with a single piezoelectric actuator, and through careful design of the resonance modes of the bristles/legs seems promising.   
\begin{figure}[t!]
    \centering
    \includegraphics[trim = 85mm 0mm 0mm 0mm, clip, width=0.45\textwidth]{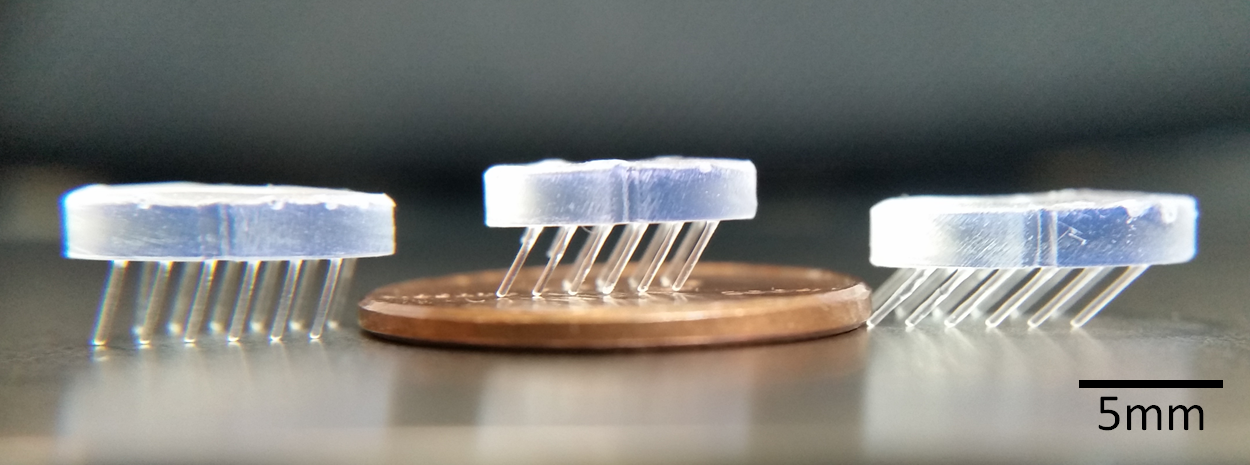}
    \caption{3D-printed bristle-bots with 60\degree~and 45\degree~leg tilt from the horizontal line, compared to a US penny.}
    \label{fig:No pzt on us penny}
\end{figure} 

The possibility of backward locomotion in bristle-bots was first predicted in \cite{desimone2012crawling} using a wet friction model, and further theoretically analyzed in~\cite{cicconofri2015motility}, and experimentally observed for a macro-scale bristle-bot (with > 10 cm feature size) using an external shaker \cite{cicconofri2016inversion}. Our work reports on the theoretical analysis of the bidirectional motion using Coulomb dry friction model, similar to the model used in \cite{Gandra2019}. Taking the analysis further, a fundamental upper limit on the robot speed as a function of actuation frequency is derived in this work for the first time. Furthermore, we extend the analysis to small-amplitude oscillations and derive the fundamental resonance frequencies in the stick and slip phases. Finally, the theoretically-predicted bidirectional motion and the speed dependence on the bristle tilt angle and actuation frequency are experimentally demonstrated using an on-board piezoelectric actuator, attached to our 3D-printed milli-bristle-bots.
Figure \ref{fig:No pzt on us penny} shows two 3D-printed bristle bots with dimensions of $12 \times 8 \times 5 \mbox{ mm}^3$ and bristle horizontal tilt angles of 60\degree~and 45\degree.

\section{Theoretical Analysis}
In this section, the equations of motion (EOMs) for the ideal motion of bristle-bots,
corresponding to a perfect slip phase following a perfect stick phase are
presented. The resonant frequencies as functions of the system parameters
for the stick and slip modes are derived in the linearized
regime, corresponding to small-amplitude oscillations.

Next, the EoMs in the presence of dry friction forces
are presented in the Cartesian coordinate system. For small-amplitude
oscillations the EoMs are simplified to piece-wise
linear equations, and the resonant frequencies, in each of these
regimes, are derived as functions of the system parameters. Starting
with a reasonable assumption on the property of the solutions of the EoMs, corroborated later by the numerical simulations, a fundamental upper-bound for the average speed of the sinusoidally
vertically driven bristle-bots is presented. 

Finally, using numerical techniques, it is verified that the EOMs in the
presence of dry friction allow for both forward and backward motion. Various phase portraits for the two directions of the motion, as well as some other interesting limits of the system (such as no kinetic friction) are provided. 

% we will see that in the case of a single 
% It is also shown that contrary to wet friction models, the EOMs of a
% dry friction model predict a minimal oscillation amplitude at a given
% frequency, below which no horizontal translational motion takes place.

% Finally all the above concepts are tied to our experimental results on
% bristle micro-bots. The experimentally measured frequency and
% amplitude ranges corresponding to different modes of motion as well as
% motion onset limits are presented and discussed for the particular
% system of micro-bots. It is observed that for a particular robot, the
% change in the drive frequency can result in a change of the direction
% and magnitude of the speed. Furthremore, a single drive function (at a
% certain amplitude and frequency) can result in forward or backward
% motion depending on the leg parameters of the robot. 

\subsection{Ideal Motion}
\begin{figure}
    \centering
    \includegraphics[trim = 25mm 0mm 25mm 0mm, clip,width=0.35\textwidth]{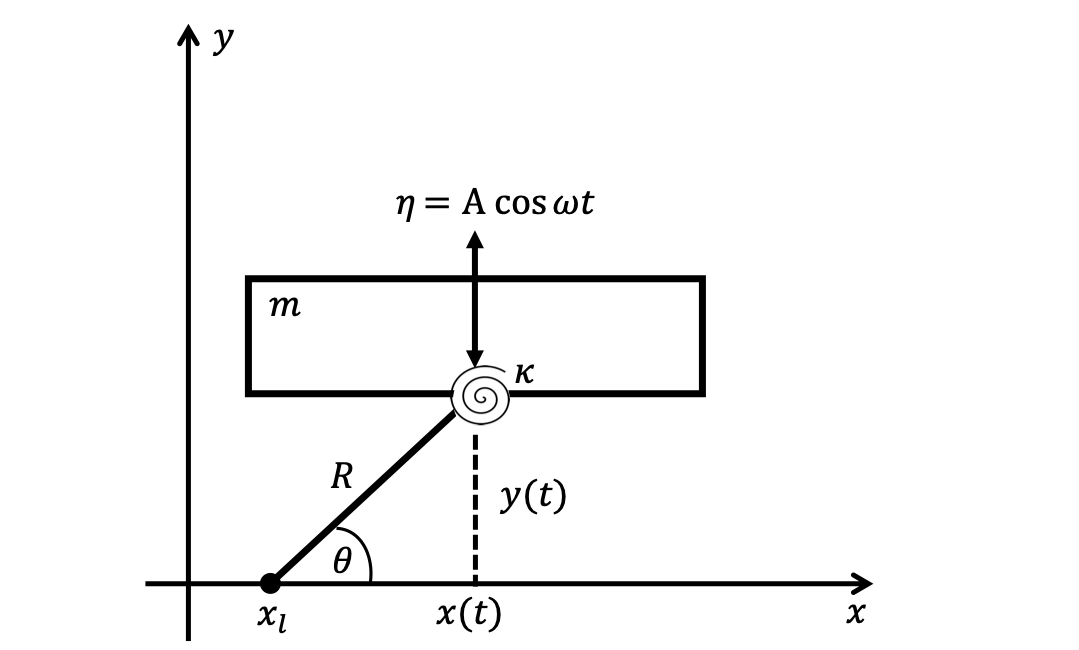}
    \caption{Depiction of the system coordinates and parameters}
    \label{fig:system_params}
\end{figure}
The ideal motion of a bristle-bot is characterized by a sequence of
alternate perfect stick and perfect slip motions. Following the
Lagrangian formalism (see the supplementary material) in the absence
of energy dissipation and external driving forces, the EoM
in the perfect stick phase is:
\begin{equation}
  \label{eq:inf_fric_eom}
  m R^2 \ddot{\theta} + m g R \cos{\theta} + \kappa (\theta - \theta_0) = 0
\end{equation},
where $m$ is the robot mass, $g$ is the gravity constant, $R$ is the
leg length, $\kappa$ is the equivalent torsional spring constant,
$\theta$ is the angle of the leg with respect to the ground
(i.e. vertical leg has $\theta = \pi/2$), and $\theta_0$ is leg angle
under no load. The system coordinates and the various parameters are depicted in the figure~\ref{fig:system_params}.

This equation is only valid in a regime in which
$ 0 \le \theta \le {\theta_0} $. The body hits the ground when
$\theta = 0$ and the legs leave the ground (robot jumps) when
$\theta = \theta_0$ and $\dot{\theta} > 0$, because the normal force
cannot be negative.

In a similar fashion, the EoMs in the perfect slip
phase, in the absence of energy dissipation and external driving
forces can be obtained by Lagrangian formalism (see supplementary
material) as:
\begin{subequations}
    \begin{equation}
    \label{eq:eom_x_no_fric}
    m \ddot{x} = 0
  \end{equation}
  \begin{equation}
    \label{eq:eom_y_no_fric}
    m\ddot{y} + m g +{ \kappa(\sin^{-1}({y/R}) - \theta_0) \over R \sqrt{1 - (y/R)^2}} = 0
  \end{equation}
\end{subequations}, where $x$ and $y$ are the horizontal and vertical
coordinates of the leg joint. So we have $\sin \theta = y/R$.

The equilibrium points $\bar{\theta}$ and $\bar{y}$ corresponding to
when the sum of forces on the robot is zero can be obtained by solving
the steady state equations (setting time derivatives to zero):
\begin{subequations}
\begin{equation}
  \label{eq:inf_fric_theta_bar}
  m g R \cos{\bar{\theta}} + \kappa (\bar{\theta} - \theta_0) = 0
\end{equation}
\begin{equation}
  \label{eq:ybar}
  m g +{ \kappa(\sin^{-1}({\bar{y}/R}) - \theta_0) \over R \sqrt{1 - (\bar{y}/R)^2}} = 0.
\end{equation}
\end{subequations}
Note that $\sin{\bar{\theta}} = \bar{y}/R$, which means $\bar{\theta}$ and
$\bar{y}$ correspond to the same equilibrium point.

One can express the equations of motion as deviations around the
equilibrium point by writing $y = \bar{y} + \hat{y}$, and
$\theta = \bar{\theta} + \hat{\theta}$. Taylor expanding and keeping
the linear terms (assuming that motions around the equilibrium
points are small), we obtain the following equations:
\begin{equation}
  \label{eq:linearized_ideal_eom}
\begin{cases}
  \ddot{\hat{\theta}} + \omega_\theta^2 \hat{\theta} = 0 & \mbox{perfect stick} \\
  \ddot{\hat{y}} + \omega_y^2\hat{y} = 0 & \mbox{perfect slip} \\
  \ddot{x}  = 0 & \mbox{perfect slip}
\end{cases}
\end{equation},
where $\omega_\theta$ and $\omega_y$ are the resonant frequencies at
perfect stick and perfect slip stages, respectively, defined as:
\begin{subequations}
  \begin{equation}
    \label{eq:omega_theta_2}
    \omega_\theta^2 \equiv \biggl({\kappa \over m R^2}\biggr)\biggl(1 + (\bar{\theta} - \theta_0) \tan{\bar \theta}\biggr)
  \end{equation}
    \begin{equation}
    \label{eq:omega_y_2}
    \omega_y^2 \equiv \biggl({\kappa \over m R^2 \cos^2{\bar \theta}}\biggr)\biggl(1 + (\bar{\theta} - \theta_0) \tan{\bar \theta}\biggr)
  \end{equation}
It can be seen that $\omega_y = {\omega_\theta / \cos \bar{\theta}}$.
\end{subequations}

Note that the ideal motion is agnostic to the direction of motion,
being backward or forward. In the forward motion the perfect stick
phase is when $\dot{\theta} < 0$, while in the backward motion the
perfect stick phase is when $\dot{\theta} > 0$. The former corresponds
to the robot pushing itself and the latter to the robot pulling
itself. In other words in the forward motion the slip phase is when
the robot is going up and in the backward motion the slip phase is
when the robot is going down.

\subsection{Dry Friction}
\label{sec:dry_friction}
In this section we relax the assumptions and write the EoMs assuming Coulomb's dry friction. The source of friction is leg
tip surface contact, denoted by the coordinate
$x_l(t) = x(t) - \sqrt{R^2 - y(t)^2}$. We will see that once we
introduce dry friction into the model, we will end up with three
regimes governed by three sets of EoMs. Two of those
regimes correspond to the kinetic friction case, corresponding to the
two cases of the leg tip moving forward and the leg tip moving backward with
respect to the surface. The third regime corresponds to the case in
which the leg tip is stationary.

The kinetic friction force can be written as:
\begin{equation}
  \label{eq:friction}
  f_{k} = - \mu_k m(\ddot{y} + g) sign(\dot{x_l}) 
\end{equation}, where $\mu_k$ is the kinetic friction coefficient. $m(\ddot{y} + g)$ is of course the normal force. Following the generalized Lagrangian formalism (see the supplementary material), the equations of motion when the leg tip is moving ($\dot{x_l} \ne 0$), i.e. when kinetic friction is in play, are as follows:
\begin{subequations}
   \begin{equation}
     \label{eq:eom_x_with_fric}
     \ddot{x}  =   - \mu_k (\ddot{y} + g) sgn(\dot{x}_l)
   \end{equation}
  \begin{equation}
  \begin{split}
    \label{eq:full_y_eom}
    (\ddot{y} + g) \biggl(1 + \mu_k sgn(\dot{x}_l) {y \over \sqrt{R^2 - y^2}}\biggr)  \\
    + { \kappa(\sin^{-1}({y/R}) - \theta_0) \over m R \sqrt{1 - (y/R)^2}} = 0
    \end{split}
  \end{equation}
   \begin{equation}
     \label{eq:eom_x_l}
     \dot{x}_l = \dot{x} + {\dot{y} y \over \sqrt{R^2 - y^2}} \ne 0
   \end{equation}
 \end{subequations}

 When the leg tip is stationary (stiction phase), the static
 friction$f_s$ comes into play. In this case the old EoMs for perfect stiction will be valid, with the caveat that the
 maximum horizontal force, i.e. the static friction force, shall not
 exceed $\mu_s m (\ddot{y} + g)$, where $m (\ddot{y} + g)$ is
 the normal force. Subsequently we have:
 \begin{subequations}
\begin{equation}
  \label{eq:static_fric_eom}
  m R^2 \ddot{\theta} + m g R \cos{\theta} + \kappa (\theta - \theta_0) = 0
\end{equation}
\begin{equation}
  \label{eq:x_of_theta)}
  x(t) = x_l + R \cos{(\theta(t))}
\end{equation}
\begin{equation}
  \label{eq:y_of_theta)}
  y(t) = R \sin{(\theta(t))}
\end{equation}, with the condition:
\begin{equation}
  \label{static_friction_conditoin}
  |\ddot{x}| < \mu_s (\ddot{y} + g)
\end{equation}
\end{subequations}

The system goes out of the slip mode into the stick mode when the
system reaches a point in the phase space for which the inequality
(\ref{eq:eom_x_l}) is invalid. Similarly, the system goes out of the
stick mode into slip mode when $|\ddot{x}| = \mu_s (\ddot{y} + g)$. At
this point if $\ddot{x} > 0$ then $\dot{x_l} < 0$ , i.e., the leg
slips backward. Conversely if $\ddot{x} < 0$, then $\dot{x_l} > 0$,
i.e. the leg slips forward.

Note that in the slip mode ($\dot{x_l} \ne 0$), once $y$ is solved,
$x$ can be obtained trivially, and in the stick mode once $\theta$ is
solved for, $x$ and $y$ can be obtained trivially. This stems from our
choice of coordinates and remains true both in the full case and in
the linearized case (see section~\ref{sec:small-amp}).

Introducing a driving force (corresponding to vertically shaking the
ground or using an on-board piezoelectric actuator) into these EoMs is trivial. One needs to note that the vertical driving
force corresponds to modulating $g$ in the surface coordinate
system. Therefore one can make the transformation
$g \to G(t) \equiv g + \ddot{\eta}(t)$, in the above equations, where
$\eta(t)$ is the kinematic equation for the surface vertical
motion. Equivalently $\ddot{\eta}(t)$ can be thought as the
\emph{downward} force pressing the robot to the floor, in the case of
an on-board actuator.

Due to the existence of the stiction when the leg tip is stationary,
one needs to overcome the force of static friction, i.e.
$ |\ddot{x}| = \mu_s (\ddot{y} + g)$, for any slipping to take place
and the robot to have any horizontal translational motion. Otherwise,
the robot will wobble with the legs stuck to the ground. This is in
contrast to the wet friction $\sim \mu \dot{x} N$ model used by
\cite{desimone2012crawling}, where the leg is stationary in the no stiction case.

\subsubsection{Small-amplitude oscillations}
\label{sec:small-amp}
We can simplify the above systems of equations, assuming small
amplitudes of oscillations around the equilibrium point. Taylor
expanding around the point and keeping the linear terms, one gets:

\begin{subequations}
  \begin{equation}
    \label{eq:x_lin_kin_fric}
    \begin{cases}
      \ddot{x} = -\mu_k[\ddot{y} + g + \ddot{\eta}(t)] & \dot{x_l} > 0\\
      x = R \cos{\theta} + x_l & \dot{x_l} = 0 \\
      \ddot{x} = +\mu_k[\ddot{y} + g + \ddot{\eta}(t)] & \dot{x_l} < 0
    \end{cases}
  \end{equation}
  \begin{equation}
    \label{eq:y_lin_kin_fric}
    \begin{cases}
      \ddot{y}  = - \omega_{y_p}^2 (y - {\bar{y}}_{p}) - \ddot{\eta}(t)  & \dot{x_l} > 0\\
      y = R \sin{\theta} & \dot{x_l} = 0\\
      \ddot{y}  = - \omega_{y_n}^2 (y - {\bar{y}}_{n}) - \ddot{\eta}(t)  & \dot{x_l} < 0
    \end{cases}
  \end{equation},
  \begin{equation}
  \begin{split}
    \label{eq:theta_lin_stick_driven}
    \ddot{\theta} = - \biggl({\kappa \over m R^2} - {\sin{\bar{\theta}}[g + \ddot{\eta}(t)] \over R}\biggr)
    (\theta - \bar{\theta})\\
    - \biggl({\cos{\bar{\theta}} \over R}\biggr)\ddot{\eta}(t)
    \end{split}
  \end{equation},valid when:
  \begin{equation}
  \begin{split}
    \label{eq:static_friction_cond_another}
    |R \ddot{\theta}  \sin{\theta} + {\dot{\theta}}^2 R \cos{\theta} |
    < \\
    \mu_s |R \ddot{\theta}  \cos{\theta} - R {\dot{\theta}}^2 \sin{\theta} + g + \ddot{\eta}(t)|
    \end{split}
  \end{equation}
  
\begin{equation}
    \label{eq:omega_y_plus}
  % \omega_{y_p} \equiv   {\omega_y^2 + {\mu_k} (\omega_y^2 \tan{\bar{\theta}}_p -
  %     {\bar{\theta}_p - \theta_0 \over R^2 m \cos^2{\bar{\theta}}_p}) \over
  %     {(1 + \mu_k \tan{\bar{\theta}_p})}^2}
    \omega^2_{y_p} \equiv 
    \kappa\left[
    \frac{
        \splitfrac{
                1 + \Delta_{\theta_p} \tan {\bar{\theta}_{p}}+
                \mu_k \biggl(\tan{\bar{\theta}}_p
                }
                {+ \Delta_{\theta_p} (\tan^2{\bar{\theta}_{p}} - {\sin{\bar{\theta}_{p}} \over \cos^2 {\bar{\theta}_{p}}} - 1)\biggr)}
        }
        {R^2 m \cos^2{\bar{\theta}}_p {(1 + \mu_k\tan{\bar{\theta}}_p)}^2}
    \right]
\end{equation}
\begin{equation}
    \label{eq:omega_y_minus}
    \omega^2_{y_n} \equiv 
    \kappa\left[
    \frac{
        \splitfrac{
        1 + \Delta_{\theta_n} \tan {\bar{\theta}_{n}}
        -\mu_k \biggl(\tan{\bar{\theta}}_n
        }
        { + \Delta_{\theta_n} (\tan^2{\bar{\theta}_{n}} - {\sin{\bar{\theta}_{n}} \over \cos^2 {\bar{\theta}_{n}}} - 1)\biggr)}}
        { R^2 m \cos^2{\bar{\theta}}_n {(1 - \mu_k \tan{\bar{\theta}}_n)}^2}
    \right]
\end{equation},
\begin{equation}
  \label{eq:delta_p}
  \Delta_{\theta_p} \equiv {\bar{\theta}}_p - \theta_0 = \biggl({-g m R \over \kappa}\biggr)\bigl(\cos{\bar{\theta}}_p + \mu_k \sin {\bar{\theta}}_p\bigr)
\end{equation}
\begin{equation}
  \label{eq:delta_n}
    \Delta_{\theta_n} \equiv {\bar{\theta}}_n - \theta_0 = \biggl({-g m R \over \kappa}\biggr)\bigl(\cos{\bar{\theta}}_n - \mu_k \sin {\bar{\theta}}_n\bigr)
\end{equation}
\end{subequations}, where ${\bar{\theta}}_{p}$ is the angle
corresponding to ${\bar{y}}_{p}$, the steady state solution of
equation \ref{eq:full_y_eom}, for the case $\dot{x_l} > 0$, and
${\bar{\theta}}_{n}$ is the angle corresponding to ${\bar{y}}_{n}$,
the steady state solution of equation \ref{eq:full_y_eom}, for the
case $\dot{x_l} < 0$. $\bar\theta$ is the solution to equation
\ref{eq:inf_fric_theta_bar}. Finally, $\eta(t)$ as mentioned earlier is
the kinematic vertical ground shake function, or equivalently the
on-board PZT shakers drives the system with the \emph{downward} force
of $\ddot{\eta}(t)$. As we can see the EOMs have been reduced to
piece-wise linear differential equations

Remarks: Note that the introduction of kinetic friction $\mu_k$ has resulted in
the bifurcation of the resonant frequencies in the slip mode
($\omega_{y_p}$ and $\omega_{y_n}$) , as well as the equilibrium
points (${\bar{\theta}}_p$ and ${\bar{\theta}}_n$).

Assuming that $\Delta_\theta \ll 1$, meaning that the legs are fairly
stiff, we can neglect the effect of $\Delta_\theta$ in frequency
bifurcation and make the following approximations:
\begin{subequations}
  \begin{equation}
    \label{eq:omega_yp_approx}
    \omega^2_{y_p} \approx {\omega^2_y \over {1 + \mu_k \tan{\bar \theta}}}
  \end{equation}
    \begin{equation}
    \label{eq:omega_yn_approx}
    \omega^2_{y_n} \approx {\omega^2_y \over {1 - \mu_k \tan{\bar \theta}}}
  \end{equation}
\end{subequations}
, where $\omega^2_y$ is given by the equation \ref{eq:omega_y_2}. This
shows that for stiff legs and small amplitude of oscillation (which is
in fact dictated by stiffness if the robot is to always remain on the
ground), $\omega^2_{y_n} > \omega^2_{y_p}$. One can also see from
eq.~\ref{eq:delta_p} and eq.~\ref{eq:delta_n}, that
\begin{equation}
  \label{eq:theta_p_gt_theta_n}
\bar{\theta}_p > \bar{\theta} > \bar{\theta}_n
\end{equation}

Next, note that the asymptotic solution of the robot motion (once all
the transient solutions die out) cannot be exclusively in the regime
$\dot{x}_l > 0$, or exclusively in the regime $\dot{x}_l < 0$. In
other words one cannot have $\dot{x}_l(t) > 0, \forall t > T$ . We
demonstrate this by employing proof by contradiction: Assume (without
loss of generality) that the steady state is purely in the regime in
which $\dot{x}_l > 0$. Then, on the one hand, by definition of
periodic steady state, we have:
\begin{subequations}
\begin{equation}
  \label{eq:periodic_steady_state_x}
  \langle \ddot x \rangle  = 0
\end{equation}
\begin{equation}
  \label{eq:periodic_steady_state_y}
  \langle \ddot y \rangle  = 0.
\end{equation}
\end{subequations}
On the other hand, by being purely in the $\dot{x}_l > 0$ regime and from the EoMs, we have:
(\ref{eq:x_lin_kin_fric}):
\begin{equation}
  \label{eq:periodic_steady_state_xl_gt_0}
  \langle \ddot x \rangle = - \mu_k \langle \ddot{y} \rangle
  - \mu_k \langle g \rangle  - \mu_k \langle \ddot{\eta}(t) \rangle  =  -\mu g  
\end{equation}. Eq.~\ref{eq:periodic_steady_state_x} is incompatible with
eq.~\ref{eq:periodic_steady_state_xl_gt_0}, so the steady state cannot by purely in this regime.

The intuitive statement of this fact is that the robot cannot steadily
walk forward in a fashion in which the leg tip is never going
backwards or at least remaining stationary. Otherwise the friction
will only act in a single direction and thus slows the robot down.

\subsubsection{Upper bound for the average speed}
Assuming a force term $\ddot{\eta}(t) \sim \sin (\omega t + \phi)$ and
enough energy dissipation in the system, we can conjecture (which is
verified later using numerical analysis) that in the asymptotic
solution, once the transients die out, $\dot{y}$ will only be zero
twice during each cycle to the first approximation. In this case the
dominant angular frequency will be the driving angular frequency
$\omega$ as expected. This means that the maximum angular range
covered in each cycle is $2 \times \theta_0$. Knowing that the leg
stop at in each cycle and the angle must recover, then the maximum
distance that the center of mass can travel in each cycle is
$R (1 - \cos\theta_0)$ corresponding to perfect stiction. This means that
the maximum average speed for the asymptotic solution is:
\begin{equation}
  \label{eq:max_avg_speed}
  \langle \dot x \rangle_{max} = \bigl({\omega \over {2 \pi}}\bigr) R (1 - \cos \theta_0)
\end{equation},
where $\omega$ is the angular frequency of the actuator, $R$ is the robot leg length and $\theta_0$ is the free leg angle.

\subsection{Numerical Solutions in Different Limits}
\label{sec:numerical}
In this section we explore the numerical solutions of the above system
of equations for different system parameters in different regimes.
All equations are solved numerically using GNU Octave\cite{octave2015},
with Livermore solver for ordinary differential equations (LSODE).

\subsubsection{Forward and backward motion}
\begin{figure*}
  \centering
  \begin{subfigure}[b]{\linewidth}
    \centering
    \includegraphics[trim = 10mm 10mm 10mm 10mm,clip,width=0.9\linewidth]{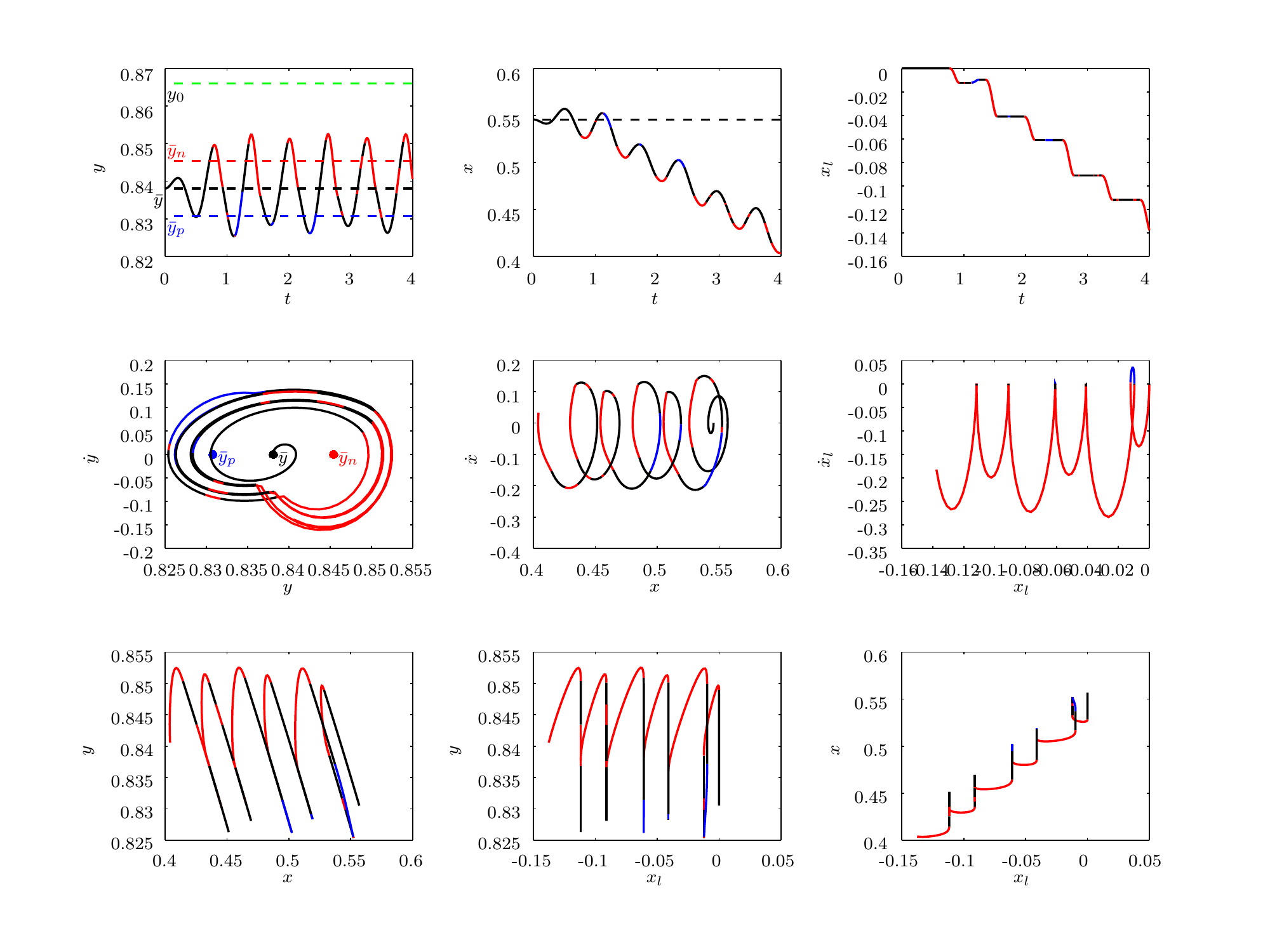}
    \caption{Backward motion: $\eta(t) = -A \cos 10 t$.}
    \label{fig:backward}
  \end{subfigure}
  \hfill
  \begin{subfigure}[b]{\linewidth}
    \centering
    \includegraphics[trim = 10mm 10mm 10mm 2mm,clip,width=0.9\linewidth]{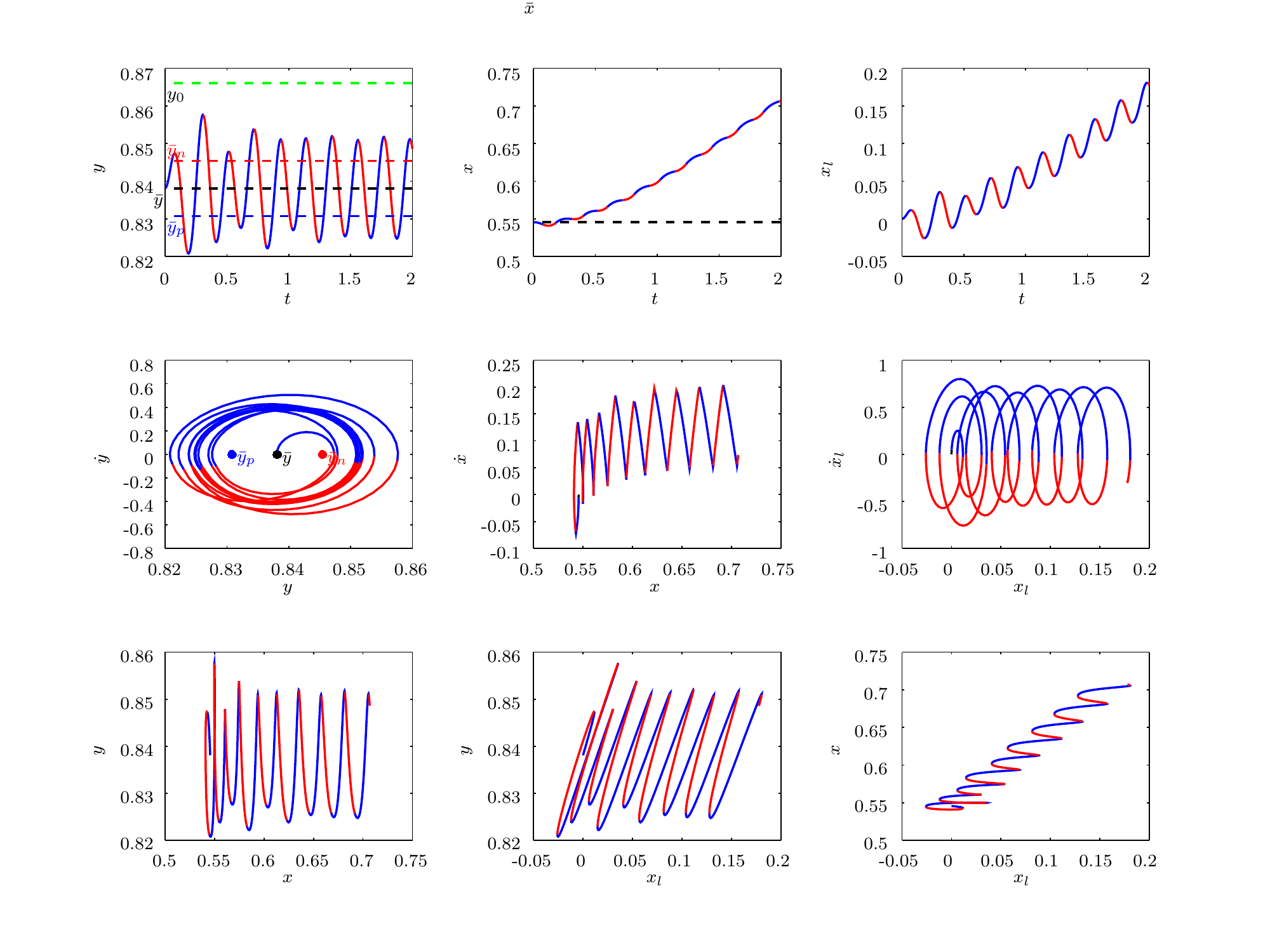}
    \caption{Forward motion: $\eta(t) = -A \cos 30 t$}
    \label{fig:forward}
  \end{subfigure}
  \caption{Various plots and phase portraits pertaining to (a) backward
    and (b) forward motion. The two systems only differ in their
    driving frequency. The system parameters for these particular
    solution are: $\kappa = 100$Nm/rad, $m = 1$Kg, $g = 9.8$ms$^{-2}$,
    $R = 1$m, $\mu_s = 0.17$, $\mu_k = 0.15$, $\theta_0 = \pi/3$,
    $A = 0.01$m, where time is measured in seconds.  For the backward
    case (a), it is seen that for about 0.25s the leg tip is
    stationary while the amplitude of the motion increases. Eventually
    the static friction is not strong enough to keep the leg tip in
    place and the leg starts sliding resulting in a net backward
    motion. In forward case (b) contrary to the backward case (a), the
    initial kick from the driving function is so strong that the
    system immediately goes into the slip mode. This stems from the
    fact that the maximum force amplitude is $\omega^2 A$, so
    everything else being the same, the higher the driving frequency
    the stronger the force amplitude. (See supplementary video)}
  \label{fig:backward_forwrad}
\end{figure*}

First, we verify that the system of equations allows for both forward
and backward motions depending on the system
parameters. Figure~\ref{fig:backward} depicts a backward motion
solution, and Fig.~\ref{fig:forward} depicts a forward motion
solution. In all the figures when the leg tip is stationary, the curve
is traced in black. When the leg tip is moving backward, the curve is
traced in red, and when the leg tip is moving forward the curve is
traced in blue. In this particular case, the only difference between
the two systems is the driving frequency.
\begin{figure*}
  \centering
  \begin{subfigure}[b]{\linewidth}
    \centering
    \includegraphics[trim = 10mm 10mm 10mm 10mm,clip,width=0.9\linewidth]{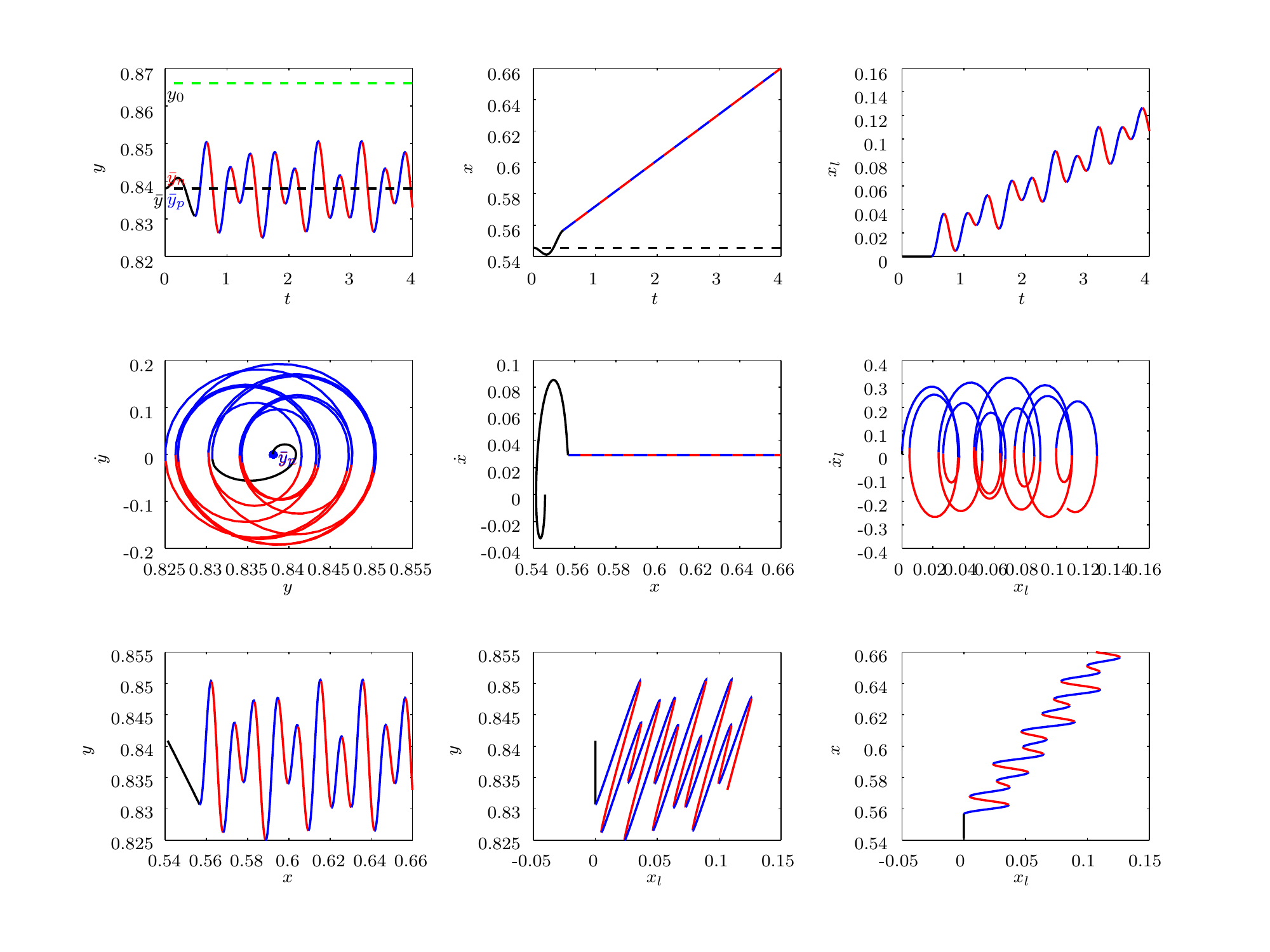}
    \caption{Forward motion: $A = 0.01$m.}
    \label{fig:nofric_forward}
  \end{subfigure}
  \begin{subfigure}[b]{\linewidth}
    \centering
    \includegraphics[trim = 10mm 10mm 10mm 2mm,clip,width=0.9\linewidth]{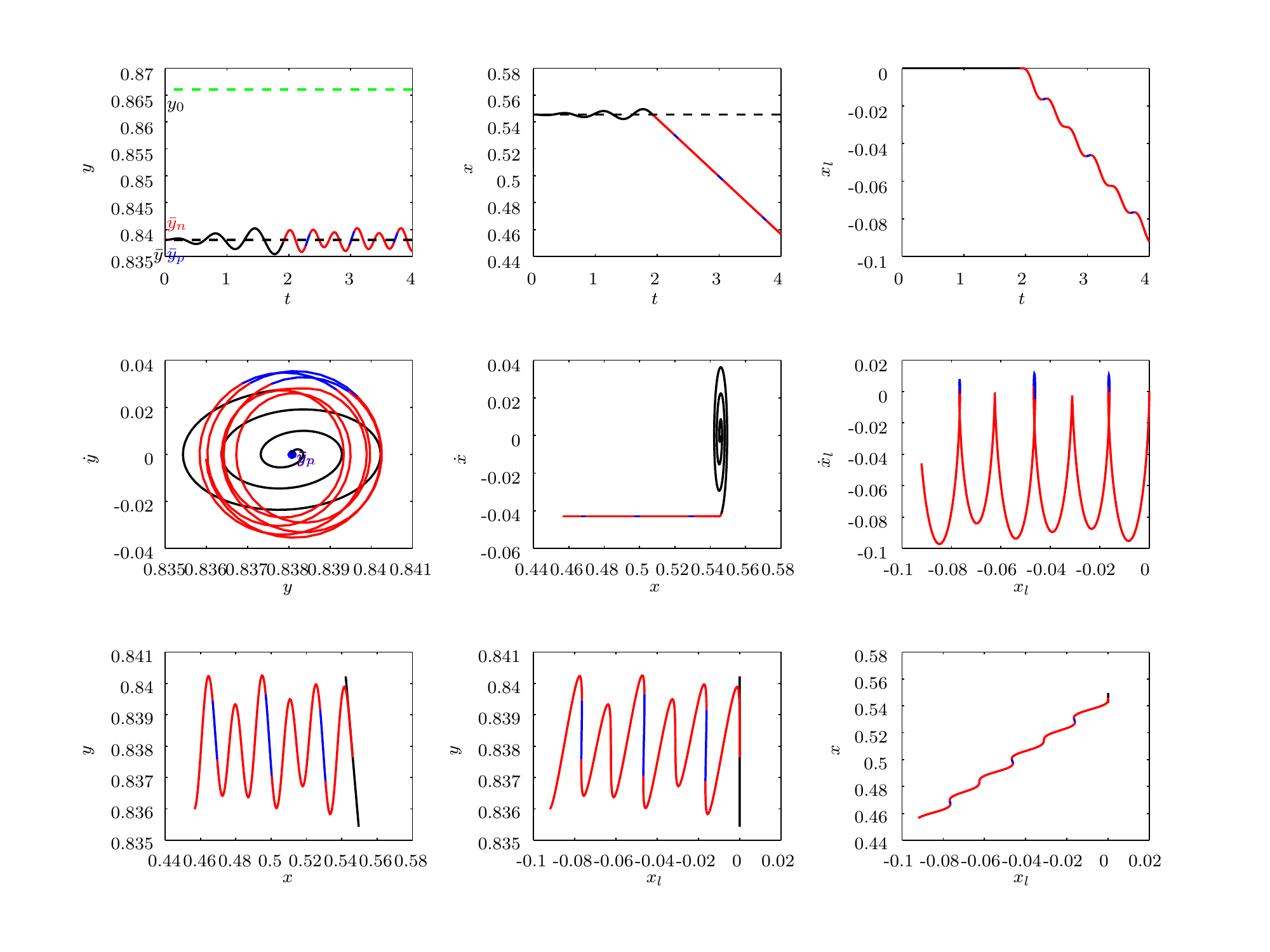}
    \caption{Backward motion: $A = 0.001$m.}
    \label{fig:nofric_backward}
  \end{subfigure}
  \caption{Forward (a) and Backward (b) motion in the case of no
    kinetic friction. The only different between the two systems is
    the driving amplitude $A$. The system parameters are
    $\kappa = 100$Nm/rad, $m = 1$Kg, $g = 9.8$ms$^{-2}$, $R = 1$m,
    $\mu_s = 0.17$, $\mu_k = 0$, $\theta_0 = \pi/3$,
    $\eta(t) = -A \cos \pi t$, (where time is measured in seconds). (See supplementary video) }
\end{figure*}
\begin{figure*}
  \centering
  \begin{subfigure}[b]{\linewidth}
    \centering
    \includegraphics[trim = 10mm 10mm 10mm 10mm,clip,width=0.9\linewidth]{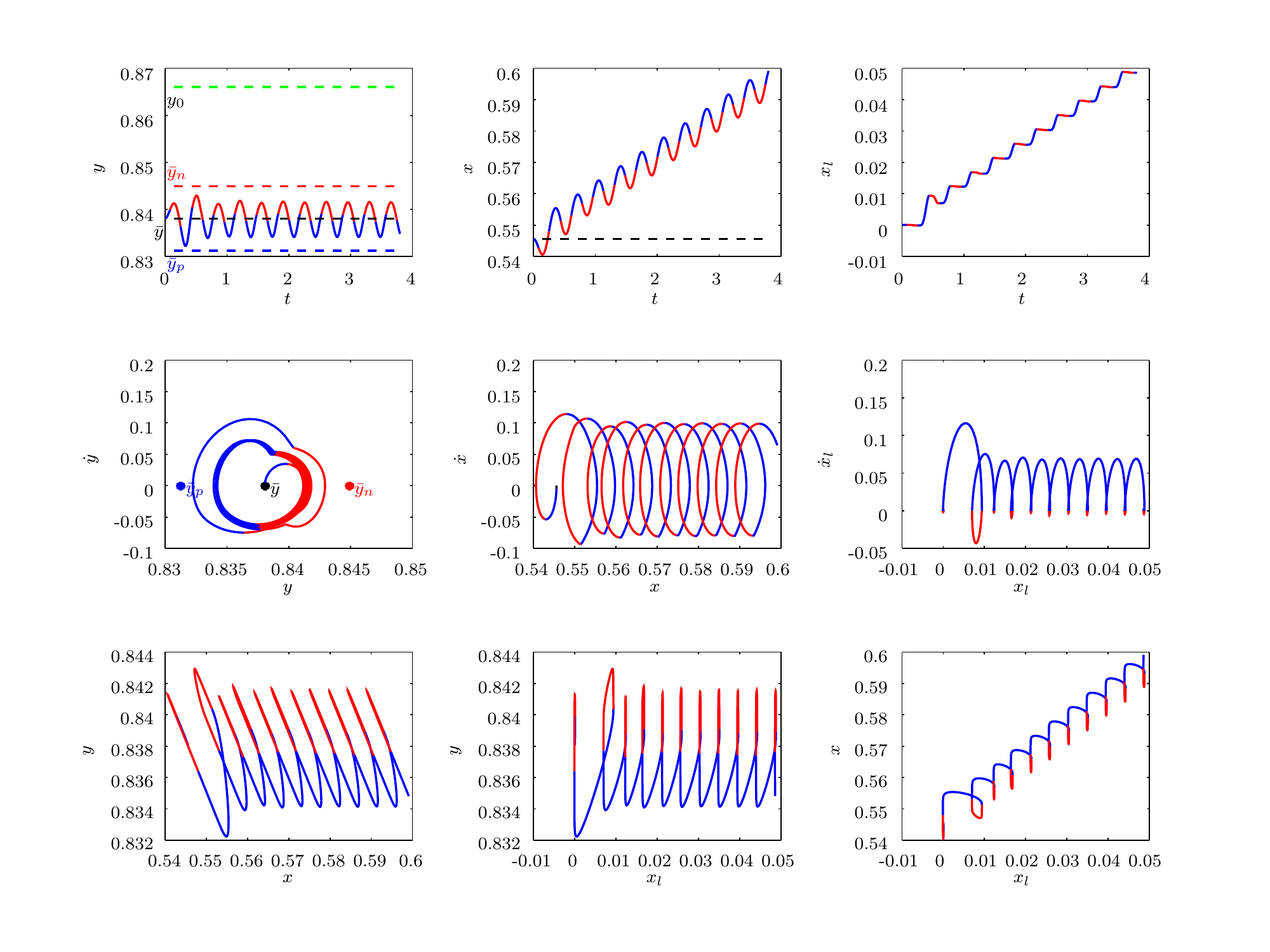}
    \caption{Non-resonant response with $\mu_k = 0.14$,
      $\omega_{y_p} = 15.77$ Hz, $\omega_{y_n} = 20.17$ Hz}
  \end{subfigure}
  \hfill
  \begin{subfigure}[b]{\linewidth}
    \centering
    \includegraphics[trim = 10mm 10mm 10mm 2mm,clip,width=0.9\linewidth]{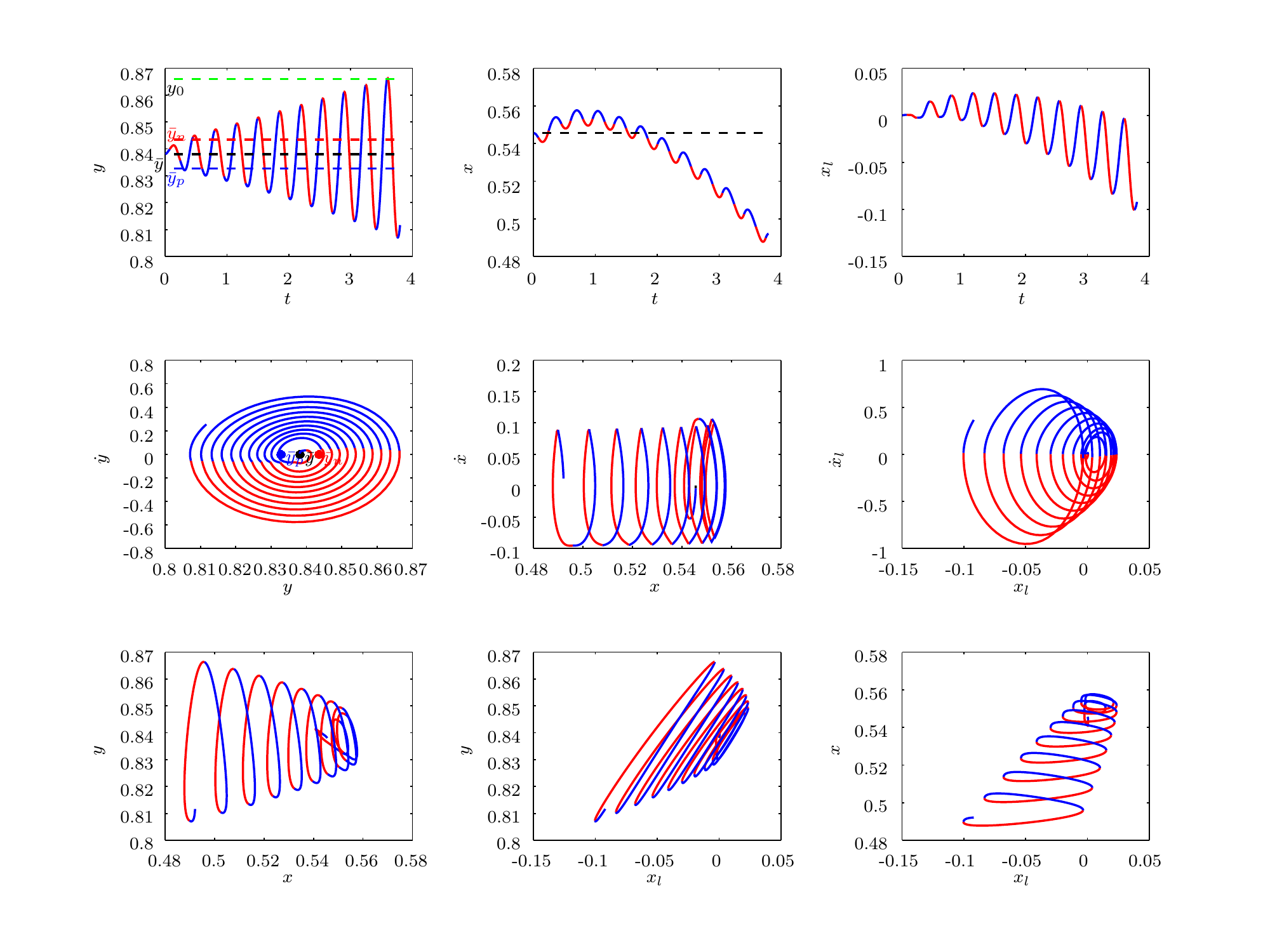}
    \caption{Resonant response with $\mu_k = 0.11$.
      $\omega_{y_p} = 16.11$ Hz, $\omega_{y_n} = 19.51$ Hz}
  \end{subfigure}
  \caption{Resonance and non-resonance behaviour for a system driven
    at a resonant frequency. The two systems share the same parameter
    values except for their $\mu_k$: $\kappa = 100$Nm/rad, $m = 1$Kg,
    $g = 9.8$ms$^{-2}$, $R = 1$m, $\mu_s = 0$, $\theta_0 = \pi/3$rad,
    $A = 0.0075$m, $\eta(t) = -A \cos 18 t$, (where time is measured in seconds), and $\omega_{y} = 17.56$. (See supplementary video)}
  \label{fig:nostatic_res}
\end{figure*}
\subsubsection{Friction, resonance, and damping }
The presence of the two kinds of friction, static and kinetic, can
lead to an interesting behavior. For example, if the static friction is strong enough, the leg tips
remain stationary and the system wobbles in place. The same effect can
be achieved if the amplitude of the driving frequency is too
small (See supplementary video).

The limiting case of no kinetic friction, but non-zero static friction
can also lead to interesting behavior. In this case by simply changing
the driving amplitude one can shoot the robot forward or
backward, in a manner similar to a sling. The final direction of the robot in this case is dependent
on the phase of motion, at the moment the static friction yields. Figures~\ref{fig:nofric_forward} and \ref{fig:nofric_backward}
exemplify such behavior. The only difference between the two is the
amplitude of the driving function.

At the other extreme, consider the artificial case in which
$\mu_s = 0$ and $\mu_k >0$. This is essentially the friction model
used by~\cite{cicconofri2016inversion}. We saw in the previous section that the
introduction of $\mu_k$ results in the bifurcation of resonance
frequencies, and the equilibrium points, one for when the leg is
slipping forward and one for when the leg is slipping backwards. It is
observed that for a system driven at $\omega$, where
$\omega_{y_p} < \omega < \omega_{y_n}$, if $\bar{y}_p < y < \bar{y}_n$
at all times, then the system does \emph{not} exhibit resonance
behavior. On the other if $y$ crosses $\bar{y}_p$ or $\bar{y}_n$
while being driven at $\omega$, where
$\omega_{y_p} < \omega < \omega_{y_n}$, then the system exhibits
resonance behavior. Making $y$ cross the limit, can be accomplished
by increasing the drive amplitude. On the other hand we can narrow the
$\bar{y}_n$--$\bar{y}_p$ range by reducing
$\mu_k$. Figure~\ref{fig:nostatic_res} exemplifies the situation. Of
course, the amplitude cannot grow indefinitely. One way for the growth
to end is hitting the $y_0$ limit, when the robot will start to jump. Even
before that, some internal damping mechanism will kick in. The damping is kept
out of the equations for simplicity. It can be introduced
if needed as ${1\over 2} \zeta \dot{\theta}^2$ dissipation function in
the Lagrangian, where $\zeta$ is the angular damping coefficient in
the stiction regime and given the relationship $y = R \sin \theta$,
it will show up as the term $\zeta \dot{y} / (R^2 - y^2)$, in the
slipping regime equation.

\section{Experimental Setup}

%\subsection{Fabrication and test setup}

\begin{figure}[h!]
    \centering
    \includegraphics[width=0.45\textwidth]{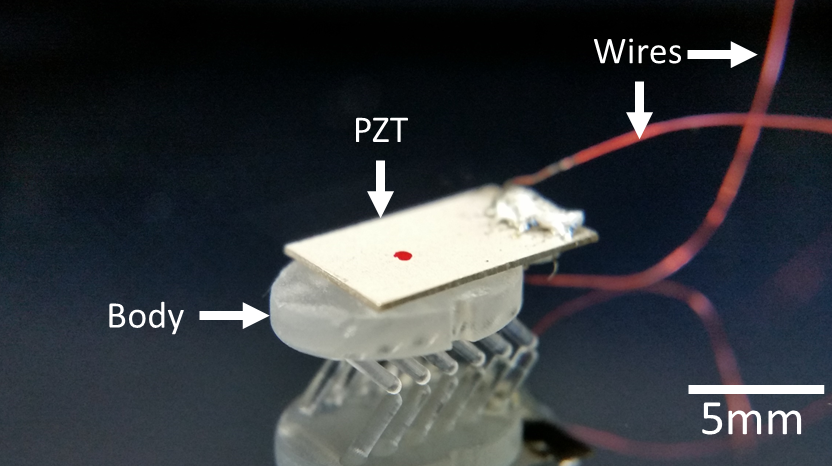}
    \caption{Assembled bristle-bot with on-board PZT actuator and soldered wires for power.}
    \label{fig:assembled_robot}
\end{figure}

The body and bristles of the bristle-bots with an overall dimension of $10\times 8\times 5\mbox{ mm}^3$ were constructed using 3D CAD software and 3D printed by Formlabs Form 3 printer, using a resin with a Young’s modulus of 2.8 GPa and a density of $1.02$ g/cm$^3$ \cite{FormlabsResin}. A rectangular PZT block of $10\times 6\times 0.3$ mm$^3$ size and a $d_{33}$ piezoelectric coefficient of $400 \times 10^{-12}$ m/V, supplied by APC International, Ltd \cite{PiezoCeramicsAPC}, is attached to the top of the 3D printed bristle-bots with adhesive, and used as the on-board source of vertical oscillatory vibrations. The on-board PZT actuator is connected with 42 gauge wires to the Krohn-Hite 7602M amplifier, which is in turn, connected to the Agilent 33220A function generator. A sinusoidal voltage input is fed to the on-board PZT actuator. A fully assembled bristle-bot is shown in Fig. \ref{fig:assembled_robot}. A dot was painted on top of the bristle-bot as a marker, and Photron FASTCAM SA3 was used to capture the motion of the bristle-bot at 60 fps with a $1280\times720$ resolution. DLTdv digitizing tool from \cite{hedrick2008software} is used for analyzing the motion by tracking the painted dot on each robot. The speed of travel is extracted by measuring the distance traveled divided by the time between two frames. A complete list of mechanical and material properties of bristle-bots are shown in Table \ref{tab:properties}. The bristle-bots were individually actuated on a glass substrate with sinusoidal signal with frequencies ranging from 10 kHz to 100 kHz in 1 kHz steps with input voltage amplitudes of 75 V. Static friction coefficient ($\mu_s$) of 0.36 and kinetic friction coefficient ($\mu_k$) of 0.32 between the robot legs and the glass substrate have been obtained experimentally by measuring the angle of slope at which the robots start sliding, and the acceleration of sliding, respectively.

%\begin{figure}[h!]
%    \centering
%    \includegraphics[width=0.45\textwidth]{figs/VideoCapture.png}
%    \caption{Setup used to record speed of bristle-bot.}
%    \label{fig:video_capture}
%\end{figure}

\begin{table}[h!]
\centering
\begin{tabular}{|l|c|}
\hline
\rowcolor{blue!25}
\begin{tabular}[c]{@{}l@{}}Property\end{tabular} & Value \\ \hline
\begin{tabular}[c]{@{}l@{}}Resin Density ($\rho$)\end{tabular} & 1.02 g/cm$^3$ \\ \hline 
\begin{tabular}[c]{@{}l@{}}Total Mass ($m$)\end{tabular} & 0.27 g  
\\ \hline     
\begin{tabular}[c]{@{}l@{}}Young's Modulus ($E$)\end{tabular} & 2.8 GPa
\\ \hline
\begin{tabular}[c]{@{}l@{}}Poisson Ratio ($\nu$)\end{tabular} & 0.35     
\\ \hline
\begin{tabular}[c]{@{}l@{}}Number of Legs\end{tabular} & 12           \\ \hline
\begin{tabular}[c]{@{}l@{}}Leg Diameter\end{tabular} & 0.8 mm
\\ \hline
\begin{tabular}[c]{@{}l@{}}Leg Length (R)\end{tabular} & 2.7 mm
\\ \hline
\begin{tabular}[c]{@{}l@{}}Leg Tilt Angle ($\theta_0$)\end{tabular} & 60\degree, 45\degree 
\\ \hline
\begin{tabular}[c]{@{}l@{}}Static Friction Coefficient ($\mu_s$)\end{tabular} & 0.36
\\ \hline
\begin{tabular}[c]{@{}l@{}}Kinetic Friction Coefficient ($\mu_k$)\end{tabular} & 0.32
\\ \hline
\end{tabular}
\caption{Mechanical and geometrical properties of the fabricated milli-bristle-bots.} 
\label{tab:properties}
\end{table}

\section{Discussions}
\subsection{Prediction of speed vs. frequency}
\begin{figure}[h!]
  \centering
  \includegraphics[trim = 10mm 0mm 10mm 0mm,clip,width=\linewidth]{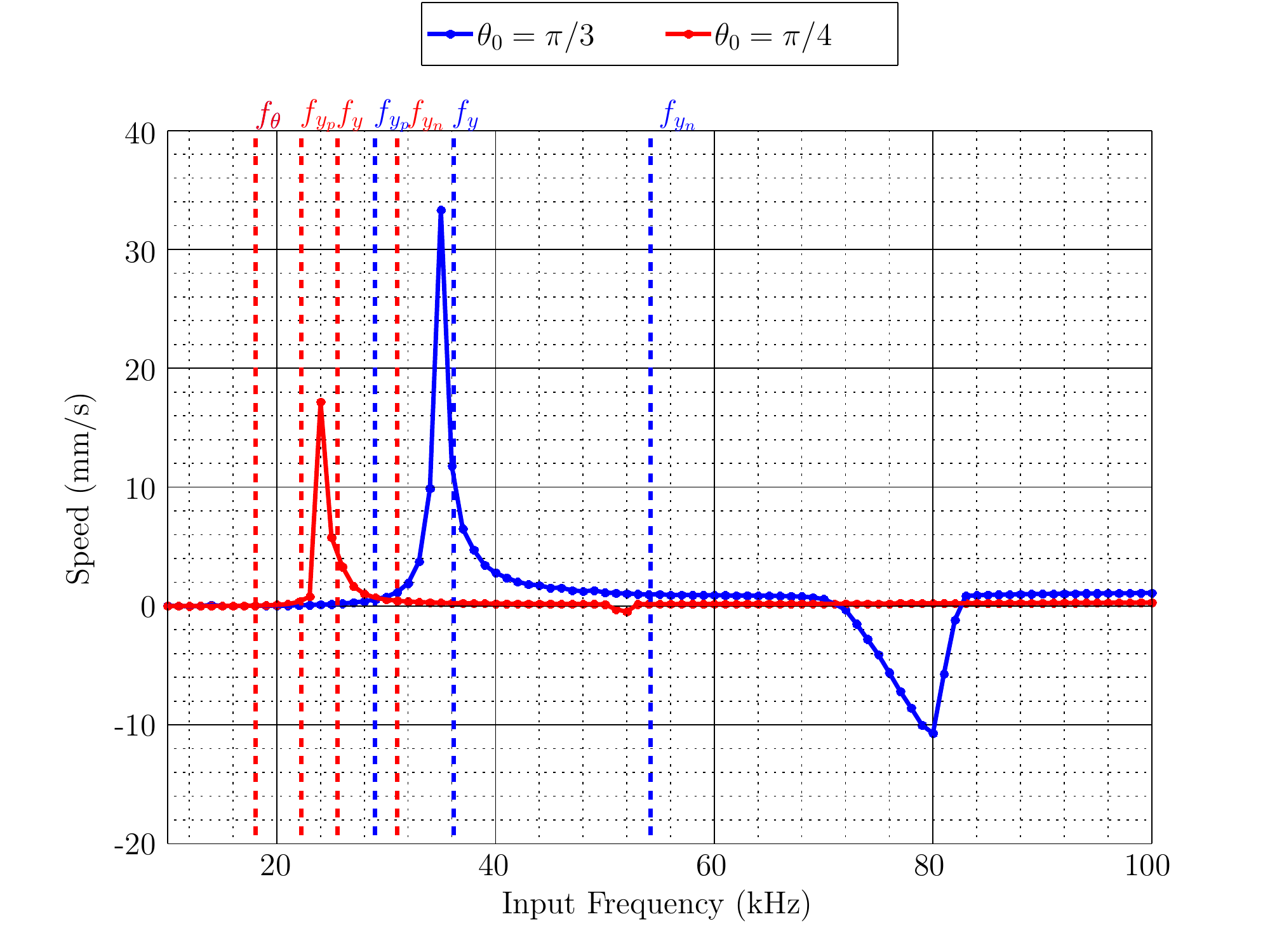}
  \caption{Speed vs. frequency, for the frequency range
    $10 \le f \le 100$ kHz with A = $10^{-8}$m, and typical robot
    system parameters at two different bristle angles from the
    horizontal line. The fundamental frequencies of the system in different
    regimes are shown for each case. Note that $f_\theta$ of the two
    systems are almost identical.}
  \label{fig:sweep_theta_pi_over_3_damp_0}
\end{figure}

In this section, we plug in the parameters from our
experimental setup into the EoMs discussed in section~\ref{sec:dry_friction}. In addition to the values for $R$, $g = 9.8$ m/s$^2$, $\theta_0$ ($\pi/3$ or $\pi/4$), $\mu_s$, $\mu_k$ and $m$ provided in Table~\ref{tab:properties}, the equations require the value for $\kappa$, and the form for $\eta$. We estimate the effective $\kappa \approx 0.1$ Nm/rad (see the supplementary material). For the driving function we choose the typical values $A = 10^{-8}$m and $\eta = A \cos (\omega t)$. 
We perform a frequency sweep (varying $\omega$ from $2\pi \times 10^4$ rad/s to $2\pi \times 10^5$ rad/s in increments of $2\pi$ rad/s), measuring the robot speed 
vs. the input frequency for the two cases of leg tilt angles. The simulation for each driving frequency ran for 2 ms (i.e. 20 cycles for the lowest frequency and 200 cycles for the highest frequency). The speed was measured by dividing the distance travelled in the last 1 ms by the time interval (i.e. 1 ms). Figure \ref{fig:sweep_theta_pi_over_3_damp_0},
depicts the speed vs. frequency response. As shown in Fig. 6, two prominent
peaks are observed, corresponding to the (1) forward and (2)
backward motions. The resonance frequencies derived for small
amplitudes in different regimes are depicted in the diagram. The first
peak, corresponds to the frequency at which the forward
speed is at maximum, appears near $\omega_y$, which is the slip case resonant frequency. The second peak, corresponding to the backward motion, is more prominent for the $\theta_0 = \pi/3$ tilt angle case.

A careful examination of the numerical solution reveals that at
points, the solutions contain $y > R \sin \theta_0$. This typically
indicates a jump, however the maximum $y - R \sin \theta_0 value is \sim 10^{-6}$ m, which is negligible considering the surface roughness of the ground and the bristles.

\subsection{Measured speed vs. frequency}
\begin{figure}[h!]
  \centering \includegraphics[trim = 10mm 0mm 10mm
  -0.1mm,clip,width=\linewidth]{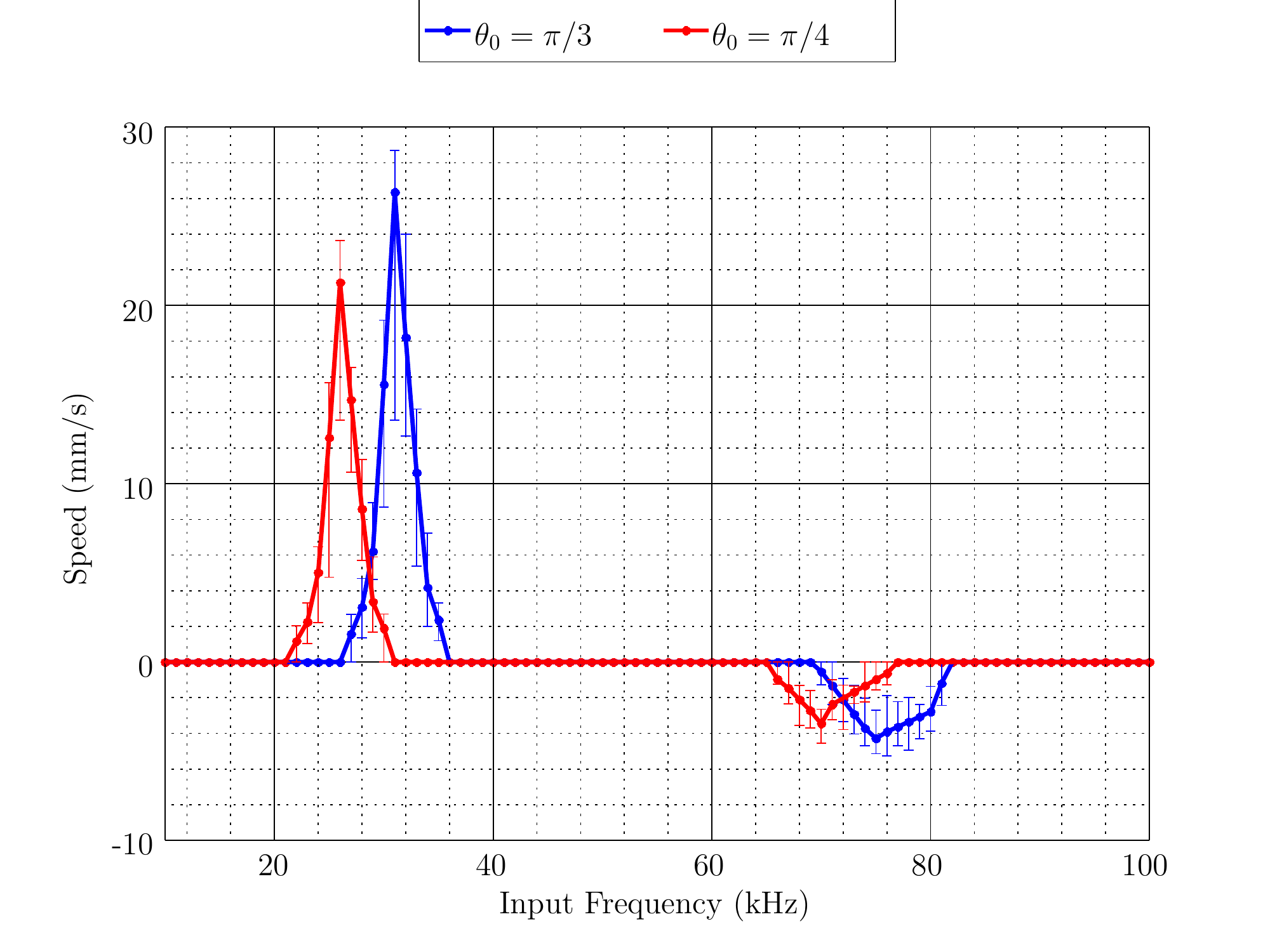}
  \caption{Experimental measurement of the speed vs. frequency of the robot with bristles having 60\degree and 45\degree~horizontal tilt angles, with the error bars shown for multiple experiment runs. The robots are actuated by a sinusoidal voltage with an amplitude of 75 V. (See supplementary video for the experimental observation of the forward and backward motions.)}
  \label{fig:experimental_sweep}
\end{figure}

A sinusoidal voltage waveform was applied to the on-board piezoelectric actuator, where the actuation frequency was swept from 10 kHz to 100 kHz, at a constant 75 V amplitude. The robot speed is plotted against actuation frequency for two different bristle tilt angles in Fig. \ref{fig:experimental_sweep}. The forward locomotion speed of the bristle-bots peaks at 27 kHz and 31 kHz, for robots with 45\degree~and 60\degree~bristle tilt angle, respectively. The backward locomotion of the corresponding bristle-bots is induced at 70 kHz and 75 kHz, respectively. The forward speed of the bristle-bots is about 5 times larger than the backward speed. The supplementary video for Fig. \ref{fig:experimental_sweep} shows the forward and backward motion of the 3D-printed milli-bristle-bot with on-board actuator.

The theoretical speed vs. frequency prediction well matches the experimentally measured values, particularly in the case of $\theta_0 = \pi/3$. Qualitatively, in both theory and experiment, there is a prominent peak around 20 to 30 kHz frequencies, corresponding to the forward motion. Then, a gap of zero or very small speed appears, followed by the backward motion region at higher frequencies, around 70 to 80 kHz, but with lower peak speeds. Note that this backward region does not correspond to the backward behavior that has been predicted or observed in \cite{Gandra2019} or \cite{cicconofri2016inversion}. Unlike this work, in the two latter cases, the backward frequency is \emph{lower} than the forward frequency and in the case of \cite{cicconofri2016inversion}, there is a well-defined transition frequency. 
For the leg angle $\theta_0 = \pi/3$, the corresponding peak frequencies for the forward and backward regions of theory and experiment are very similar. Furthermore, the theory predicts that reducing the bristle angle to $\pi/4$ reduces the resonant frequency of the purely slip case (which is proportional to $1/\cos \bar{\theta}$) and thus shifts the peak speed frequency to the left. This effect was observed in the experiment as well. The numerical solutions also suggest that the peak speed for the lower bristle angle is lower than that of the higher bristle angle, and the experimental results match the prediction. The discrepancies observed between the theory and experiment, regarding the backward regions of $\theta_0 = \pi/4$ case, are attributed to the non-idealities and assembly imperfections, such as off-centered placement of the PZT block on the robot body, or tension due to the wires applying forces on the experimental bristle bots. Furthermore, the theoretical model assumes the robot to move in 2D (i.e. locomotion in x-axis, with body oscillations in y-axis), while the experimental bristle bots move in 3D space (locomotion in x- and y-axis, with body oscillations in z-axis) which leads to greater internal frustration of the multi-legged bristle-bot system.

\section{Conclusion}
We provided a theoretical analysis of the dynamics of bristle robots, subject to dry friction and derived the fundamental frequencies of the system. We also reported on the fabrication of milli-bristle-bots with a single on-board actuator and varying bristle tilt angles. The body and bristles of the bristle-bots were 3D-printed and a PZT actuator plate was first diced and then attached to the robot body to provide on-board oscillatory vibrations. In accordance with our theoretical prediction, backward motion frequency regions were experimentally observed at higher frequencies than the forward motion frequency regions, unlike previous reports. We also reported for the first time on the effect of the bristle angles on the locomotion speed. As predicted by our theoretical model, the peak speed frequency was indeed reduced as the bristles were further tilted towards the horizontal line. 

The bidirectional motion using a single on-board actuator, reported here for the first time, is a stepping-stone for achieving full steering capability at the millimeter and micrometer scales. Further enhancements to the theory including taking non-idealities such as weight distributions and adhesion forces into account might be necessary for a more accurate prediction of the response of the bristle-bots.

%Achieving bi-directional motion using a single on-board actuator is stepping stone in fabricating 

%It also compares 
%This paper compares the theoretical and experimental results of the forward and backward motion of . As predicted by the theory two distinct frequency regions were experimentally observed, corresponding to the forward and backward locomotion. However, the conclusion can be made that experimental results support the theoretical model of bristle bot's forward and backward locomotion and that change in leg tilt angle can induce shift in actuation frequencies of bi-directional locomotion with single on-board piezoelectric actuator.

\medskip
\bibliographystyle{elsarticle-num}
\bibliography{BackwardPaper_BibtexFormatReferences}

\begin{thebibliography}{10}
\expandafter\ifx\csname url\endcsname\relax
  \def\url#1{\texttt{#1}}\fi
\expandafter\ifx\csname urlprefix\endcsname\relax\def\urlprefix{URL }\fi
\expandafter\ifx\csname href\endcsname\relax
  \def\href#1#2{#2} \def\path#1{#1}\fi

\bibitem{cicconofri2016inversion}
G.~Cicconofri, F.~Becker, G.~Noselli, A.~Desimone, K.~Zimmermann, The inversion
  of motion of bristle bots: analytical and experimental analysis, in:
  Symposium on Robot Design, Dynamics and Control, Springer, 2016, pp.
  225--232.

\bibitem{becker2014spy}
F.~Becker, S.~B{\"o}rner, T.~K{\"a}stner, V.~Lysenko, I.~Zeidis, K.~Zimmermann,
  Spy bristle bot—a vibration-driven robot for the inspection of pipelines,
  in: 58th Ilmenau Scientific Colloquium, Ilmenau, Germany, Sept, 2014, pp.
  8--12.

\bibitem{wang2008bristle}
Z.~Wang, H.~Gu, A bristle-based pipeline robot for ill-constraint pipes,
  IEEE/ASME Transactions On Mechatronics 13~(3) (2008) 383--392.

\bibitem{giomi2013swarming}
L.~Giomi, N.~Hawley-Weld, L.~Mahadevan, Swarming, swirling and stasis in
  sequestered bristle-bots, Proceedings of the Royal Society A: Mathematical,
  Physical and Engineering Sciences 469~(2151) (2013) 20120637.

\bibitem{klingner2014stick}
J.~Klingner, A.~Kanakia, N.~Farrow, D.~Reishus, N.~Correll, A stick-slip
  omnidirectional powertrain for low-cost swarm robotics: mechanism,
  calibration, and control, in: 2014 IEEE/RSJ International Conference on
  Intelligent Robots and Systems, IEEE, 2014, pp. 846--851.

\bibitem{desimone2012crawling}
A.~DeSimone, A.~Tatone, Crawling motility through the analysis of model
  locomotors: two case studies, The European Physical Journal E 35~(9) (2012)
  85.

\bibitem{cicconofri2015motility}
G.~Cicconofri, A.~DeSimone, Motility of a model bristle-bot: A theoretical
  analysis, International Journal of Non-Linear Mechanics 76 (2015) 233--239.

\bibitem{ioi1999mobile}
K.~Ioi, A mobile micro-robot using centrifugal forces, in: 1999 IEEE/ASME
  International Conference on Advanced Intelligent Mechatronics (Cat. No.
  99TH8399), IEEE, 1999, pp. 736--741.

\bibitem{schulke2011worm}
M.~Schulke, L.~Hartmann, C.~Behn, Worm-like locomotion systems: development of
  drives and selective anisotropic friction structures,
  Universit{\"a}tsbibliothek Ilmenau, 2011.

\bibitem{Senyutkin1977bristle}
A.~Senyutkin, Bristle bot, Available at \url{http://zhurnalko.net/
  =sam/junyj-tehnik/1977-06num53} (2015/12/14) (1977).

\bibitem{Gandra2019}
C.~Gandra, Dynamics of a vibration driven bristle bot, Ph.D. thesis, Clemson
  University (2019).

\bibitem{octave2015}
S.~H. John W.~Eaton, David~Bateman, R.~Wehbring,
  \href{http://www.gnu.org/software/octave/doc/interpreter}{{GNU Octave}
  version 4.0.0 manual: a high-level interactive language for numerical
  computations}, 2015.
\newline\urlprefix\url{http://www.gnu.org/software/octave/doc/interpreter}

\bibitem{FormlabsResin}
Formlabs, Formlabs Material Properties- Standard:Photopolymer Resin for Form 2
  3D Printers, rev. 1 (4 2017).

\bibitem{PiezoCeramicsAPC}
APC International, Ltd., Physical and Piezoelectric Properties of APC
  Materials, rev. 4 (1 2019).

\bibitem{hedrick2008software}
T.~L. Hedrick, Software techniques for two-and three-dimensional kinematic
  measurements of biological and biomimetic systems, Bioinspiration \&
  biomimetics 3~(3) (2008) 034001.

\end{thebibliography}

\end{document}

\maketitle
\section{Lagrangian and EoM for the Stick Case}
In the Lagrangian formalism $L = T - V$, where $T$ is the kinetic
energy and $V$ is the potential energy.  For a single-leg system, when
the leg tip stationary, in the absence of energy dissipation and an
external driving force, the kinetic energy ($T$) of the system can be
written as:
\begin{equation}
  \label{eq:inf_fric_T}
  T = {1 \over 2} m (R \dot{\theta})^2
\end{equation}
The potential energy ($V$) of the system can be written as:
\begin{equation}
  \label{eq:inf_fric_V}
  V =  m g ( R \sin{\theta} + Y_0) + {1 \over 2} K (\theta - \theta_0)^2
\end{equation}, where $Y_0$ is just a constant and denotes how much higher the center of mass is from the leg joint. Since this parameter has no effect on the equation of motion we shall drop it  from the future Lagrangians.

The equations of motions are thus:
\begin{equation}
  \label{eq:lagrange_eom_stick}
  {d \over dt} {\partial L \over \partial \dot{\theta}} - {\partial L
    \over \partial \theta} = 0
\end{equation}

This yields:
\begin{equation}
  \label{eq:inf_fric_eom}
  m R^2 \ddot{\theta} + m g R \cos{\theta} + K (\theta - \theta_0) = 0
\end{equation}
\section{Lagrangian and EoM for the Pure Slip Case}
For the pure slip case, in the Cartesian coordinate system, the
Lagrangian of the system can be written as:
\label{sec:no_friction}
\begin{equation}
  \label{eq:L_no_friction}
  L = {1 \over 2} m \dot{x}^2 + {1 \over 2} m \dot{y}^2 - m g y - {1 \over 2} K  (\sin^{-1}\big({y \over R}\big) - \theta_0)^2
\end{equation}
where we have written the angle ($\theta$) as $\sin^{-1}(y/R)$.

Similarly we have

\begin{subequations}
  \begin{equation}
    \label{eq:lagrange_eom_slip_x}
    {d \over dt} {\partial L \over \partial \dot{x}} - {\partial L
      \over \partial x} = 0
  \end{equation}
  \begin{equation}
    \label{eq:lagrange_eom_slip_y}
    {d \over dt} {\partial L \over \partial \dot{y}} - {\partial L
      \over \partial y} = 0
  \end{equation}
\end{subequations}
This yields:
\begin{subequations}
    \begin{equation}
    \label{eq:eom_x_no_fric}
    m \ddot{x} = 0
  \end{equation}
  \begin{equation}
    \label{eq:eom_y_no_fric}
    m\ddot{y} + m g +{ K(\sin^{-1}({y/R}) - \theta_0) \over R \sqrt{1 - (y/R)^2}} = 0
  \end{equation}
\end{subequations}

\section{Lagrangian and EoM for the Kinetic Friction Case}
We model the surface friction term in accordance with Coulomb's
friction force for dry surfaces
\begin{equation}
  \label{eq:friction}
   - \mu N (t) sign(\dot{x_l}) 
 \end{equation}
 Where $N$ is the force normal to the surface and $\dot{x}_l$ is the
 velocity of the leg surface contact.

 Noting that in the absence of a driving force $m\ddot{y} = N - m
 g$. (note that when driving force is present $g \to g - \ddot{\eta}$)
 The friction force in the x direction can be written as
\begin{equation}
  f_x = -\mu m (\ddot{y} + g) sgn(\dot{x}_l) 
\end{equation}
The non conservative force of the friction can hence be written as:
\begin{subequations}
  \begin{equation}
    \label{eq:Q_x}
    Q_x = f_x {\partial {x_l} \over \partial {x}} = f_x
  \end{equation}
  \begin{equation}
    \label{eq:Q_y}
    Q_y = f_x {\partial {x_l} \over \partial {y}} = {y f_x \over \sqrt{R^2 -(y)^2}} 
  \end{equation}
\end{subequations}
Noting that $x_l = x - \sqrt{R^2 - y^2}$, the new equations can be
written as
\begin{subequations}
   \begin{equation}
     \label{eq:eom_x_with_fric}
     \ddot{x}  =   - \mu (\ddot{y} + g) sgn(\dot{x}_l)
   \end{equation}
  \begin{equation}
    \label{eq:full_y_eom}
    (\ddot{y} + g) \biggl(1 + \mu sgn(\dot{x}_l) {y \over \sqrt{R^2 - y^2}}\biggr)  + { K(\sin^{-1}({y/R}) - \theta_0) \over m R \sqrt{1 - (y/R)^2}} = 0
  \end{equation}
   \begin{equation}
     \label{eq:eom_x_l}
     \dot{x}_l = \dot{x} + {\dot{y} y \over \sqrt{R^2 - y^2}}
   \end{equation}
 \end{subequations}
 
 \section{Deriving $\kappa$, Torsional Spring constant}
Our fabricated bristle-robots have 12 legs, therefore, the equivalent $\kappa$
(torsion spring constant) is nominally $12$ times
that of a single leg spring constant.
The bending spring constant $k_b$ of a single leg can be calculated
as:
\begin{equation}
  \label{eq:k_b}
  k_b = 3 E I / R^3
\end{equation}
where, $E$ is the is Young's modulus, $I$ is the second moment of the
area, and $R$ is the leg length.  Assuming that the deflection is
small, i.e., $\delta x \ll R$ , which is the case in our
setup, the change of angle is $\delta \theta \approx \delta x /
R$. From the definition of $\kappa \equiv \tau / \delta \theta$, where
$\tau$ is the torque, we have:
\begin{eqnarray*}
  \kappa \times \delta \theta &= & \tau \\
  \kappa \times \delta x / R  & = & F \times R \\
  \kappa  & = & {F \over \delta x }\times  R^2 \\
  \kappa & = & k_b \times R^2\\
\end{eqnarray*}

Plugging the numbers for our system, $k_b = 5115 N/m$ for a single leg  which results in an equivalent $\kappa = 12 \times k_b \times R^2 = 1.79 \times 10^{-1}$
Nm/rad for the overall system. However, in practice and upon closer inspection of the robot legs, not all 12 of the
legs touch the ground or bear the same weight, stemming from manufacturing 3D non-idealities. The effective
contribution of all legs is more like six or seven times that of a
single leg. The experimental data can be used to fit for this coefficient. But as we will see our initial estimate resulted in %adequate fit.  Alternatively one
can consider $\kappa$ as a free parameter and derive the value
experimentally. At any rate we used the effective $\kappa \sim 0.1$Nm/rad.   

%Finally :
%\[
%  \sqrt{\kappa \over m R^2} \approx 1.14 \times 10^5
%\].